\documentclass{elsart4-1}

\usepackage{amssymb}

\usepackage[english,francais]{babel}

\newtheorem{e-proposition}[theorem]{Proposition}

\newtheorem{e-definition}[theorem]{Definition\rm}

\renewcommand{\theequation}{\arabic{equation}}
\def\og{\leavevmode\raise.3ex\hbox{$\scriptscriptstyle\langle\!\langle$~}}
\def\fg{\leavevmode\raise.3ex\hbox{~$\!\scriptscriptstyle\,\rangle\!\rangle$}}

\usepackage [dvips]{graphicx}
\usepackage{amsmath}
\usepackage{epsfig}
\usepackage{amssymb}
\newcommand{\C}{\mathcal C}

\newcommand{\WW}{\mathbb W}
\newcommand{\KS}{\text{KS}}

\newcommand{\ee}{\text{e}}

\newcommand{\bN}{\text{\bf N}}

\newcommand{\T}{\mathcal T}
\newcommand{\HH}{\mathcal H}

\begin{document}

\centerline{Physics}
\begin{frontmatter}


\selectlanguage{english}
\title{Thermodynamic formalism and large deviation functions in continuous time Markov dynamics}


\selectlanguage{english}
\author[msc]{Vivien Lecomte},
\ead{vivien.lecomte@univ-paris-diderot.fr}
\author[lpt]{C\'ecile Appert-Rolland},
\ead{Cecile.Appert-Rolland@th.u-psud.fr},
\author[msc]{Fr\'ed\'eric van Wijland}
\ead{fvw@univ-paris-diderot.fr}

\address[lpt]{Laboratoire de Physique Th\'eorique (CNRS UMR8627), B\^atiment 210,
Universit\'e Paris-Sud, 91405 Orsay cedex, France}
\address[msc]{Laboratoire Mati\`ere et Syst\`emes Complexes,
(CNRS UMR7057), Universit\'e Denis Diderot (Paris VII),
10 rue Alice Domon et L\'eonie Duquet, 75205 Paris cedex 13, France}


\medskip
\begin{center}
{\small Received *****; accepted after revision +++++}
\end{center}

\begin{abstract}
The thermodynamic formalism, which was first developed for dynamical systems
and then applied to discrete Markov processes, turns out to be well suited
for continuous time Markov processes as well, provided the definitions are
interpreted in an appropriate way. Besides, it can be reformulated
in terms of the generating function of an observable,
and then extended to other observables.
In particular, the simple observable $K$ giving the number of events
occurring over a given time interval turns out to contain already the
signature of dynamical phase transitions.

For mean-field models in equilibrium, and in the limit of large systems,
the formalism is rather simple to apply and shows how thermodynamic phase
transitions may modify the dynamical properties of the systems.
This is exemplified with the $q$-state mean-field Potts model,
for which the Ising limit $q=2$ is found to be qualitatively different
from the other cases.

\vskip 0.5\baselineskip

\selectlanguage{francais}
\noindent{\bf R\'esum\'e}

\vskip 0.5\baselineskip
\noindent
{\bf Formalisme thermodynamique et grandes d\'eviations dans les syst\`emes \`a dynamique markovienne.}

Le formalisme thermodynamique, qui a d'abord \'et\'e d\'evelopp\'e
dans le cadre des syst\`emes dynamiques puis appliqu\'e aux processus de Markov,
s'av\`ere \'egalement pertinent pour les dynamiques de Markov en temps
continu, \`a condition toutefois d'interpr\'eter les d\'efinitions en jeu
de fa\c con appropri\'ee. Ce formalisme peut \^etre reformul\'e en termes
de fonction g\'en\'eratrice d'une observable, puis \'etendu \`a d'autres
observables. En particulier, l'observable $K$ donnant le nombre d'\'ev\'enements
ayant lieu dans un intervalle de temps donn\'e, bien que tr\`es simple,
contient d\'ej\`a la signature de transitions de phases dynamiques.

Pour les mod\`eles de champ moyen \`a l'\'equilibre, et dans la limite
des grands syst\`emes, le formalisme peut s'appliquer simplement et
montre comment les transitions de phase thermodynamiques peuvent affecter
les propri\'et\'es dynamiques de ces syst\`emes.
Cela est illustr\'e sur le cas du mod\`ele de Potts en champ moyen,
et il s'av\`ere que le cas d'Ising diff\`ere qualitativement des
autres cas.

 \vskip 0.5\baselineskip
\noindent{\small{\it Mots-cl\'es~:} formalisme thermodynamique~; grandes d\'eviations~;
chaos}}
\end{abstract}
\end{frontmatter}


\selectlanguage{english}

\section{Introduction}

The theory of chaos has first been developed in the frame of
dynamical systems. In particular the well-known
Lyapunov exponents were then defined, which characterize the exponential divergence
between initially closeby trajectories in phase space.
Lyapunov exponents characterize individual directions, and
if at least one of these exponents is positive, the system
is said to be chaotic.
The Kolmogorov-Sinai entropy provides a more global measurement of chaos.
This can directly be seen from Pesin's theorem, which relates the $KS$ entropy
to the sum of positive Lyapunov exponents: $h_{\KS} = \sum_{\lambda_i>0}
\lambda_i$ for a closed system.
In practice, in order to obtain the $h_{\KS}$ entropy, one has to
partition phase space. A trajectory in phase space becomes, in
this coarse grained description, a sequence of cells which are visited
by the actual trajectory. By doing so, the deterministic dynamics
can be turned into a stochastic one.
The $\KS$ entropy is then defined as $h_{\KS} = - \lim_{t\to\infty} \frac{1}{t}  
              \sum_{\text{histories}}
 \text{Prob}\{ \text{history} \} \ln \text{Prob}\{ \text{history} \} $.
Then, in order to mimic the formalism that is used for equilibrium systems,
a so-called thermodynamic formalism was introduced \cite{ruelle78},
in which various quantities -- and in particular the $\KS$ entropy -- are derived
from a partition function 
\begin{equation}
Z(s,t) = \sum_{\text{histories }\\ \text{from }0\to t} (\text{Prob}\{ \text{history} \})^{1-s}
\end{equation}
This function is called {\em dynamical partition function}, as the sum is not taken
over configurations but over histories.
We have replaced the usual exponent $\beta$ by $1-s$ for reasons that will become
obvious later. Anyhow, this exponent has nothing to do with a temperature.
It is just a parameter that allows to scan the distribution $\text{Prob}\{ \text{history} \}$.
In the same way as a free energy is defined for equilibrium systems,
one introduces the so-called topological pressure, or Ruelle pressure,
defined by
\begin{equation}
\psi(s) = \lim_{t\to\infty} \frac{1}{t} \ln Z(s,t)
\end{equation}
where the thermodynamic limit is replaced by a long time limit.
The $\KS$ entropy is immediately obtained from $\psi(s)$ via the relation
$h_\KS = \psi'(0)$. Other quantities that can be obtained from $\psi$ are
for example the topological entropy,
$h_\text{top} = \psi(1)$, which counts the number of trajectories
within a certain time interval, or the escape rate (for open systems)
which is equal to $\gamma = - \psi(0)$.

As the deterministic dynamics had to be turned into a stochastic dynamics
in order to apply this formalism, the latter is of course particularly well-suited
for systems which are stochastic from the beginning, as Markov processes
are.
The formalism was successfully applied to a number of Markov processes in discrete time,
and in this frame, it was shown that $h_\KS$ could be efficiently calculated
via the formulation
$h_\KS = -\frac{1}{\tau}\langle\sum_{\C'}w(\C\to\C') \ln w(\C\to\C') 
\rangle_\text{st}$ where $w(\C\to\C')$ is the probability to jump from
a configuration $\C$ to a configuration $\C^\prime$, and $\tau$ is the
time step of the process \cite{gaspard04}.

If one tries to extend the calculations of $h_\KS$ to continuous time
by letting the time step $\tau$ tend to zero, the probability $w(\C\to\C')$
(which is proportional to $\tau$) also tends to zero and $h_\KS$ diverges.
In the same way, a finite continuous time limit for the thermodynamic formalism cannot be 
obtained by taking $\tau \to 0$.

A first remark is that this divergence of $h_\KS$ does reflect the fact
that the dynamical complexity
of the trajectories is infinite when continuous time dynamics is considered, since one
needs an infinite amount of information to describe the continuum of time intervals 
between configuration changes (see Sec.\ref{sec:thermodynform}).
In that sense, the divergence is meaningful.
On the other hand, this divergence indicates that the glasses one is wearing
to look at Markov processes (discrete time)
may not be the most appropriate ones if one is interested in
Markov dynamics (i.e. with continuous time).

In this paper, we shall summarize which glasses should be used, i.e.
how the thermodynamic formalism should
be applied in the case of Markov dynamics in order to get finite quantities~\cite{lecomte_a_v05,lecomte_a_v06}.
The formalism will be rephrased in terms of large deviations of observables
- or more precisely in terms of cumulant generating functions, which are
the Legendre transforms of large deviation functions.
This allows to cast the thermodynamic formalism
into a more general frame including the much-discussed
fluctuation theorem.

The computation of the topological pressure $\psi(s)$ for physically relevant
models is in general a difficult task.
Here, we shall show that in the special case of 
mean-field models at equilibrium, the dominant term of $\psi(s)$
in the limit of large systems can be obtained quite simply from
the minimization of a given functional.
As a result, one obtains a description of the dynamical
properties of the aforementionned equilibrium systems. 
This will be illustrated on the example of the Potts model.

\section{The thermodynamic formalism for continuous time Markov dynamics}
\label{sec:thermodynform}

We shall start with some obvious remarks on the origin of
stochasticity of the trajectories.
In order to define a trajectory, one has to determine:
\begin{itemize}
\item which configurations will be visited:
this is the configurational part of the trajectory, and
it is just a sequence of configurations $\C_0\to\ldots\to\C_K$ where
successive configurations are different.
Once we know that the system will jump from configuration
$\C$ to another one, the probability to chose a specific
configuration $\C^\prime$ is $\frac{W(\C \to \C^\prime)}{r(\C)}$
where $r(\C)$ is the rate of escape from configuration $\C$:
\begin{equation}
r(\C)=\sum_{\C'\neq\C}W(\C\to \C').
\end{equation}
The probability of the configurational part of the trajectory
thus reads:
\begin{equation}
\text{Prob}\{\text{history}\}=\prod_{n=0}^{K-1} 
\frac{W(\C_n\to\C_{n+1})}{r(\C_n)}
\end{equation}
\item when shall the system jump from one configuration to the
next one:

\begin{equation}
\begin{array}{cccccccccc}
0& & t_1 & & t_2 & & \cdots & t_K & & t\\
& \C_0 & \Longrightarrow &  \C_1  & \Longrightarrow &  \C_2&  \cdots & \Longrightarrow &  \C_K 
\end{array}
\end{equation}
The probability density for not leaving the configuration $\C_{n-1}$ during the time interval $t_n-t_{n-1}$ and for changing configuration at time $t_n$  is
$ r(\C_{n-1})\ee^{(t_n-t_{n-1})r(\C_{n-1})}$.
For the last time interval, one must only ensure that the system does not leave
configuration $\C_K$ during the time interval $t-t_{K}$, and the associated
probability density is $ \ee^{(t-t_K)r(\C_K)}$.
\end{itemize}

The key point when applying the definition for the dynamical partition
function is that only the probability of the configurational part of the 
trajectory should be raised to the power $1-s$, i.e. 
\begin{eqnarray}
 Z(s,t|\C_0,t_0) & = & \sum_{k=0}^{+\infty} \sum_{\C_1,\ldots,\C_k}
  \int_{t_0}^t dt_1\:
 r(\C_0)e^{-(t_1-t_0)r(\C_0)}  \ldots \nonumber \\
& & 
   \int_{t_{k-1}}^t dt_k\:
 r(\C_{k-1})e^{(t_k-t_{k-1})r(\C_{k-1})}
 e^{-(t-t_k)r(\C_k)}
 \left[
  \prod_{n=1}^k  \frac{W(\C_{n-1}\rightarrow\C_{n})}{r(\C_{n-1})} 
 \right]^{1-s}
\label{defZ}
\end{eqnarray}
At this stage, it may seem quite arbitrary. The justification for this point of view
is mainly that it gives a coherent picture, and that
the value of $Z(s,t)$ and derived quantities can effectively be computed in several
models.  Actually, the idea was already implicitly present in a work
by van Beijeren, Dorfman and Latz
\cite{vanbeijeren_d02,latz_b_d96}, in the case of a Lorentz gas
with discs scatterers, and a hard sphere gas.

In order to give a unified picture, it is useful to notice that
$Z(s,t)$ can be expressed in terms of an observable
\begin{equation}
Q_+ = \sum_{n=0}^{K-1}\ln \frac{W(\C_n\to\C_{n+1})}{r(\C_n)}
\end{equation}
Indeed one has $Z(s,t) = \langle e^{-s Q_+}\rangle$ and the dynamical partition function appears as the moment generating
function of $Q_+$. Then $\psi(s)$ is the corresponding cumulant generating function.
The average sign $\langle ... \rangle$ includes the average both
on configurations and times.
From now on, $\psi$ will be denoted by
$\psi_+(s)$ to recall that it is associated to
the observable $Q_+$.

In practice, $Z(s,t)$ is rarely evaluated directly from its definition
(\ref{defZ}).
One rather evaluates directly $\psi(s)$, which turns out to be
the largest eigenvalue of an operator
$\WW_+(\C,\C')=W(\C'\to \C)^{1-s} r(\C')^s-r(\C)\delta_{\C,\C'}$
which is a kind of evolution operator weighted according to the parameter $s$.
In \cite{lecomte_a_v06}, we found that not only the eigenvalue but also the
eigenvector associated with this operator can yield useful information
on the dynamical structure of the system dynamics, in particular when
dynamical transitions occur.

The $\KS$ entropy can then be obtained from several definitions
which can be shown to be equivalent:
\begin{eqnarray}
h_\KS & = & -\lim_{t \rightarrow \infty}\frac{1}{t}\left\langle\sum 
  P(\C_0,\C_1,\cdots,\C_K) \ln P(\C_0,\C_1,\cdots,\C_K)
\right\rangle
\label{defhks1}
\\
& = & \psi_+'(0)
\label{defhks2}
 \\
& = & - \lim_{t \rightarrow \infty}\frac{\langle Q_+ \rangle}{t} 
\label{defhks3}
\\
& = & -\left\langle\sum_{\C'} 
    W(\C\to\C') \ln \frac{W(\C\to\C')}{r(\C)}\right\rangle_{st}
\label{defhks4}
\end{eqnarray}
The last relation was obtained by a similar derivation as the one
performed for discrete time Markov processes by Gaspard \cite{gaspard04}.

In the spirit of Gaspard \cite{gaspard98},
once the thermodynamic
formalism is expressed in terms of an observable, one can generalize
it to a whole family of observables
\begin{equation}
A(t)=\sum_{n=0}^{K-1}
\alpha(\C_n,\C_{n+1})
\label{defA}
\end{equation}
which depends only on the configurational part of the trajectory.
Then $\psi_A(s)=\lim_{t\to\infty}\frac{1}{t}\ln
\langle \ee^{-s A}\rangle$ is the moment generating function
of $A$ and can be obtained as the largest eigenvalue of an operator
$\mathbb{W}_A$ defined in a way similar to $\mathbb{W}_+$.

Among all these observables, at least two other ones than $Q_+$ seem
interesting. The first one is $K$, i.e. the number of
configuration changes within the time interval $[0,t]$.
It is the simplest one one could think of, and still
already contains some relevant information on dynamical
phase transitions.

The second one was introduced by Lebowitz and Spohn \cite{lebowitz_s99}
\begin{equation}
Q_S=\sum_{n=0}^{K-1}\ln\frac{W(\C_n\to\C_{n+1})}{W(\C_{n+1}\to \C_n)}
\end{equation}
It is this observable that verifies the fluctuation relation which,
in terms of the moment generating function $\psi_S$, reads
$\psi_S(s)=\psi_S(1-s)$ (we assume through the whole paper
that the systems under consideration can take only a finite number of
states).
Moreover, in the steady state, $Q_S$ was identified as the integrated
entropy flux~\cite{lebowitz_s99} (we are now referring to the Boltzmann entropy 
$S(t) = - \sum_{\mathcal C} P(\mathcal C,t) \ln P(\mathcal C,t)$).

We have shown that, if one defines a twin observable $Q_-$ for
the reversed trajectories, namely $Q_-= \sum_{n=0}^{K-1}\ln\frac{W(
\C_{n+1}\to\C_n)}{r(\C_{n+1})}$, then the observable $Q_+$ that
we have introduced in the frame of the thermodynamic formalism
can be related to $Q_S$ via the relation
\begin{equation}
Q_S = Q_+-Q_-
\end{equation}
and, similarly to what Gaspard did for discrete time systems \cite{gaspard04},
one can define a reversed $\KS$ entropy
$h_\KS^{R}  =  -\lim_{t\to\infty}\frac{\langle Q_-\rangle}{t}$
which relates, in the stationary state, to the entropy flux:
\begin{equation}
\sigma_\text{f}=-\sigma_{irr}=h_\KS-h_\KS^{R}
\end{equation}
where $h_\KS$ is defined by (\ref{defhks1}-\ref{defhks4}).

\section{Mean-Field models}

In general, calculating the cumulant generating function $\psi_A$ associated
with the observable $A$ is a difficult task (see
\cite{lecomte_a_v06,derrida_a99} for examples).
However, for mean-field models which can be described
by a macroscopic order parameter (such as a magnetization),
and for which a detailed balance relation is available,
one can quite easily get the leading order of $\psi_A$ in the limit of large systems,
at least for $A = K$ and $A = Q_+$.

The derivation is based on the fact that for a {\em symmetric}
operator, the largest eigenvalue can be expressed as
\begin{equation}
\lambda_\text{max} = \max_V\left\{
\frac{\langle V | \mathbb{W}^\text{sym} | V\rangle}{\langle V|V\rangle}\right\}
\end{equation}

\subsection{Symmetrization of the operators $\WW_K$ and $\WW_+$}

One can check, using the detailed balance relation
\begin{equation}
\frac{P_\text{eq}(\C')}{P_\text{eq}(\C)} = 
\frac{W(\C\to \C')}{W(\C'\to \C)},
\end{equation}
that the following operators obtained by a similarity transformation
from $\WW_K$ and $\WW_+$ are symmetric
\begin{eqnarray}
\mathbb W_K^\text{sym} = & \T^{-1} \mathbb W_K \T & \mbox{ ; }\;\; 
\T(\C,\C') = P_\text{eq}^{1/2}(\C)\: \delta_{\C,\C'}
\label{simil_K}\\
\mathbb W_+^\text{sym} = & \T^{-1} \mathbb W_+ \T & \mbox{ ; }\;\;
\T(\C,\C') = 
P_\text{eq}(\C)^{\frac{1-s}{2}}r(\C)^{-\frac{s}{2}}\: \delta_{{\C,\C'}}
\label{simil_plus}
\end{eqnarray}
As a property of similarity transformations, these symmetric operators
have the same spectrum - and thus the same largest eigenvalue -
as the original ones.

To make the presentation easier to follow, we shall exemplify
the remaining of the approach in the case of the Potts model.

\subsection{Definition of the mean-field Potts model}

In each site of a lattice of size $N$, a spin variable $\sigma_i$
can be in $q$ different states. The mean-field Potts model is
defined by the Hamiltonian
\begin{equation*}
\HH = -\frac{J}{2N} \sum_{i,j} \left[q \delta_{\sigma_i,\sigma_{j}}
-1 \right] = -\frac{J}{2N} \sum_{k=1}^q \left[ N_k^2
-N^2 \right]
=  -\frac{J}{2} N \left[ q \sum_{k=1}^q \theta_k^2 -1\right]
\end{equation*}
where $N_k$ is the number of spins in state $k$ and
$\theta_k = \frac{N_k}{N}$ the corresponding fraction.
When a site switches from state $n$ to state $m$, the
corresponding energy variation writes
\begin{equation}
\Delta \HH = -\frac{Jq}{N} \left[ N_m - N_n + 1 \right]
= -Jq \left[ \theta_m - \theta_n + 1/N \right]
\end{equation}
The model is endowed with a continuous time Glauber like dynamics
with transition rates
\begin{equation}
W(\sigma_i=n \to \sigma_j=m) = \ee^{-\beta \frac{\Delta \HH}{2}}
\end{equation}
The equilibrium probability distribution
\begin{equation}
P_\text{eq}\left(\left\{\sigma_i\right\}\right) = \frac{1}{Z}
\ee^{-\beta \HH}
\end{equation}
verifies detailed balance.
However, in what follows, we shall rather describe the system
at the level of the occupation numbers $\bN = \left\{N_k\right\}_{k=1..q}$. Then
the equilibrium distribution that obeys detailed balance reads
\begin{equation}
P_\text{eq}\left(\bN\right) = \frac{1}{Z}
\frac{N!}{\Pi_{k=1}^q N_k!}\ee^{-\beta \HH}
\end{equation}
with the transition rates
\begin{equation}
W\left(\bN',\bN\right) = W\left(\bN \to \bN'\right) = 
N_n \ee^{\beta \frac{Jq}{2N} \left[ N_m - N_n + 1 \right]}
\end{equation}
where $\bN' =
\left\{N_1,\cdots,N_n'=N_n-1,\cdots,N_m'=N_m+1,\cdots,N_q\right\}$.

In the following, we shall consider the operator associated with
the observable $K$:
\begin{equation}
\WW_K\left(\bN',\bN\right)
= z \sum_n \sum_{m\neq n} \delta_{N_n',N_n-1} \delta_{N_m',N_m+1}
\Pi_{k\neq n, k\neq m} \delta_{N_k',N_k}
W\left(\bN \to \bN'\right)
- \Pi_{k=1}^q \delta_{N_k',N_k} r(\bN)
\end{equation}
where $z = \ee^{-s}$.
Using the same transformation as in (\ref{simil_K}),
one obtains the symmetric operator
\begin{eqnarray*}
\WW_K^\text{sym}\left(\bN',\bN\right)
& = &z \sum_n \sum_{m\neq n} \delta_{N_n'+1,N_n} \delta_{N_m'-1,N_m}
\prod_{k\neq n, k\neq m} \delta_{N_k',N_k} \left[(N_n'+1) N_m' \right]^{1/2} \\
&  & \quad -\prod_{k=1}^q \delta_{N_k',N_k} 
 \sum_n \sum_{m\neq n} 
N_n' \ee^{\beta \frac{Jq}{2N} \left[ N_m' - N_n' + 1 \right]}
\end{eqnarray*}
For large $N$, and outside of the thermodynamic transition,
one expects that the eigenvector associated with the
largest eigenvalue will take the form
\begin{equation*}
V(\bN)\sim \ee^{N f(\left\{\theta_k\right\}_{k=1..q})}
\end{equation*}
Indeed, this is the case in particular for $z=1$ ($s=0$), i.e. for
the equilibrium distribution with
\begin{equation}
f(\left\{\theta_k\right\}_{k=1..q}) = \beta \frac{Jq}{2} \sum_{k=1}^q \theta_k^2
\end{equation}
An expansion in the large $N$ limit yields
\begin{eqnarray}
&&\sum_{\bN'} \WW^\text{sym}(\bN,\bN')Q(\bN') \nonumber\\
& = & 
z \sum_n \sum_{m\neq n} \sqrt{(N_n+1)N_m}
\ee^{N f(\theta_1,\cdots,\theta_n+\frac{1}{N},\cdots,\theta_m-\frac{1}{N},\cdots,\theta_q)}
- \sum_n \sum_{m\neq n} N_n\ee^{\beta \frac{Jq}{2N}\left[N_m-N_n+1\right]}
\ee^{N f(\theta_1,\cdots,\theta_q)} \nonumber\\
& = & N \ee^{N f(\theta_1,\cdots,\theta_q)} \sum_n \sum_{m\neq n}\left\{
z \sqrt{\theta_n \theta_m}
\ee^{\frac{\partial f}{\partial \theta_n} - \frac{\partial f}{\partial \theta_m}}
- \theta_n \ee^{\beta \frac{Jq}{2} \left[\theta_m-\theta_n\right]}
\right\}
\label{eq_dvlpt}
\end{eqnarray}
The quantity to maximize is
\begin{equation}
\frac{\sum_{\bN} \sum_{\bN'} V(\bN) \WW^\text{sym}(\bN,\bN')V(\bN')}{
\sum_{\bN} \left[V(\bN)\right]^2} 
= \frac{\sum_{\bN} N\ee^{2Nf(\theta_1,\cdots,\theta_q)} \sum_n \sum_{m\neq n}
\left\{ z\sqrt{\theta_n \theta_m}
\ee^{\frac{\partial f}{\partial \theta_n} - \frac{\partial f}{\partial \theta_m}}
- \theta_n \ee^{\beta \frac{Jq}{2} \left[\theta_m-\theta_n\right]}
\right\}
}{\sum_{\bN} \ee^{2Nf(\theta_1,\cdots,\theta_q)}}
\end{equation}
For each function $f$, in the large $N$ limit, the sum is dominated by
the maximum of $f$, for which $\frac{\partial f}{\partial \theta_n}=0$ $\forall n=1..q$, i.e.
\begin{equation}
\frac{\psi_K(s)}{N} = \max_{\bN} \left\{
\sum_n \sum_{m\neq n}\left[ z \sqrt{\theta_n \theta_m} 
- \theta_n \ee^{\beta \frac{Jq}{2} \left[\theta_m-\theta_n\right]}\right]
\right\}
\label{maxeq}
\end{equation}

We shall now apply this result to various values of $q$.

\subsection{Arbitrary value of $q$ }

As in \cite{mendes_l91}, we assume a symmetry breaking along one
particular direction $k=1$ (assume for example that a very small
field along this direction breaks the symmetry of the system).
Then the system can be described by one order parameter
$m \in [-1/(q-1),1]$ with $\theta_k = \frac{1-m}{q}$ $\forall k > 1$ and
$\theta_1 = \frac{1+(q-1)m}{q}$.

The cumulant generating function for $K$ is then obtained from
\begin{eqnarray}
\frac{\psi_K(s)}{N} = &&
\max_{m} \left\{
2 \frac{q-1}{q} z \sqrt{(1-m)\left[1+(q-1)m\right]}
+ \frac{(q-1)(q-2)}{q} (z-1) (1-m)
\right.\nonumber \\ &&\left. \quad \quad \quad
- \frac{q-1}{q}\left[1+(q-1)m\right]
\ee^{- \frac{\beta Jq}{2} m}
- \frac{q-1}{q} (1-m) \ee^{\frac{\beta Jq}{2} m}
\right\}
\label{maxeqq}
\end{eqnarray}

We shall now take special values of $q$ to explore the consequences
of this maximization condition.

\subsection{$q=2$: the mean-field Ising model}

\begin{figure}
\begin{center}
\includegraphics[height=50mm]{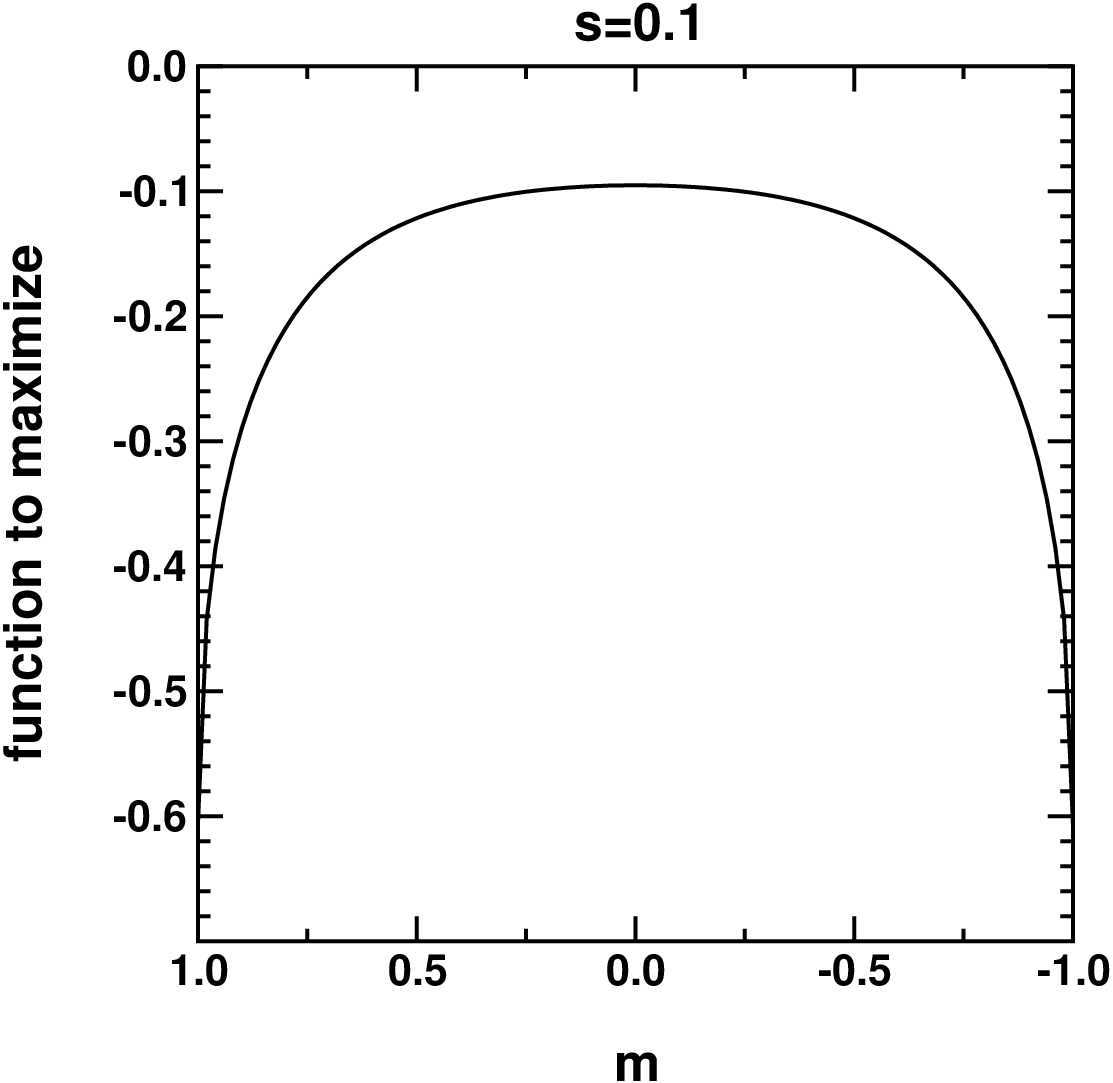}
\includegraphics[height=50mm]{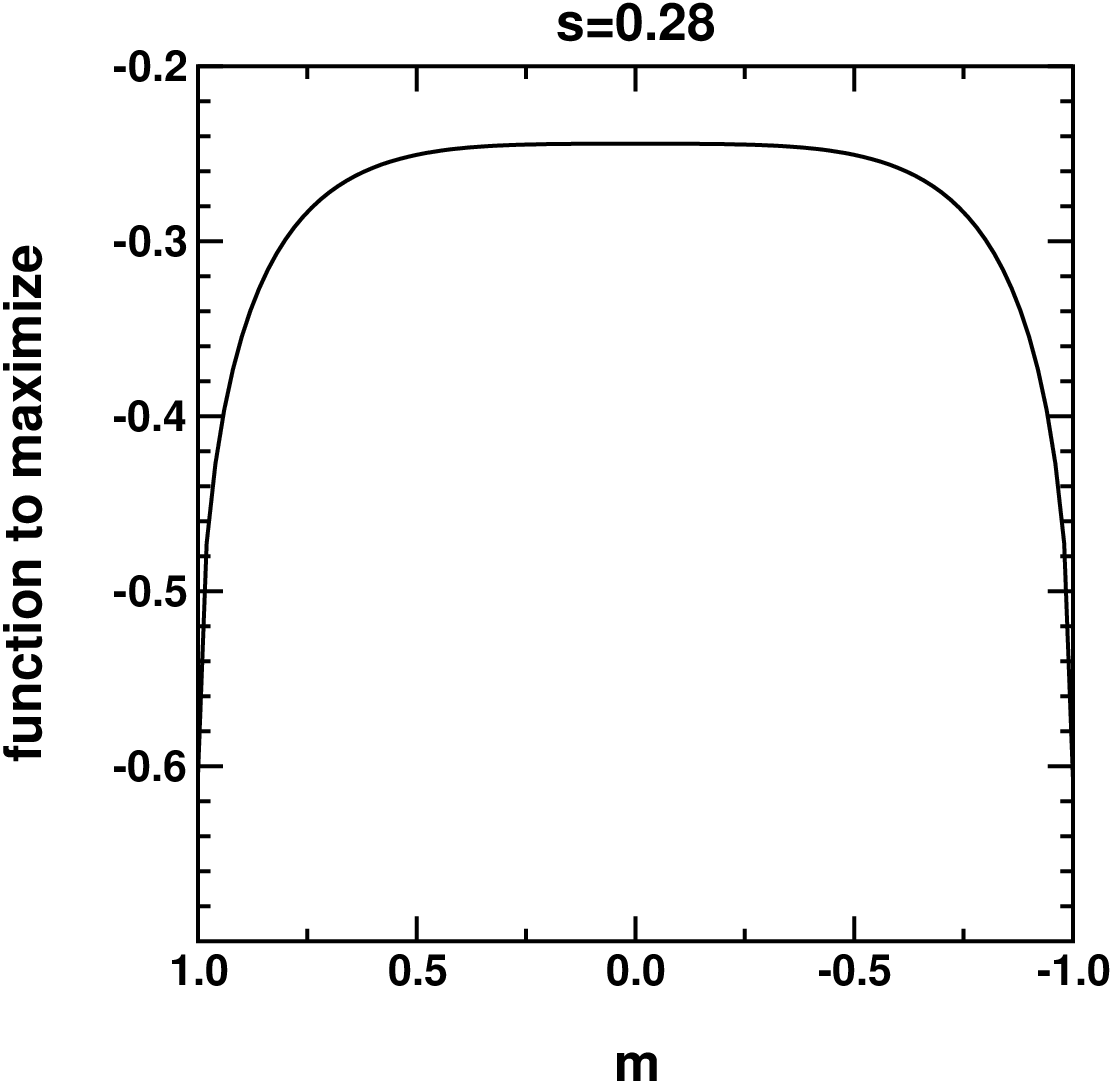}
\includegraphics[height=50mm]{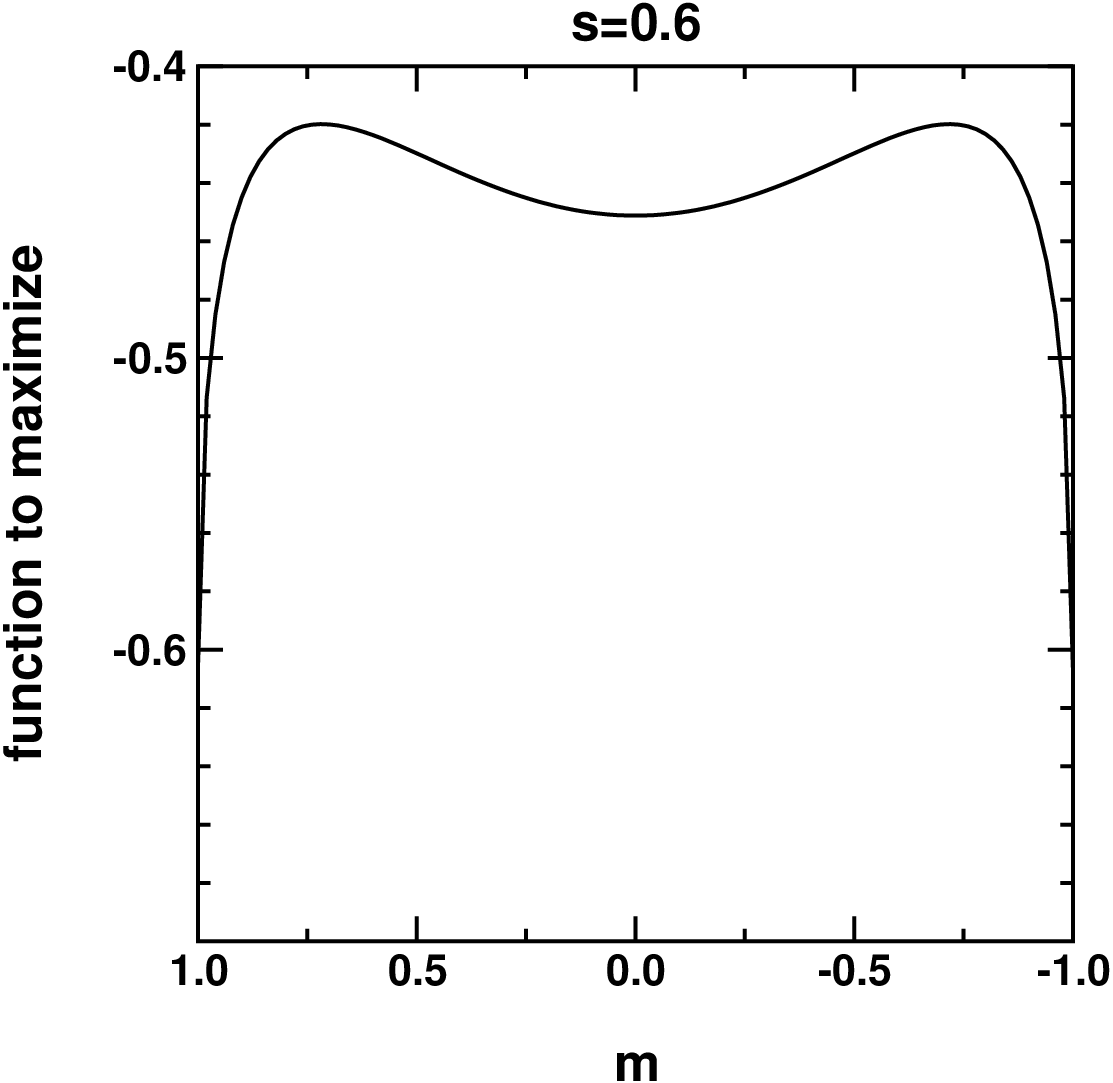}
\end{center}
\caption{The function that must be maximized in order to find
the value of $m_K(s)$ and $\frac{\psi_K(s)}{N}$, in the case
of the Ising model (q=2), for a temperature $\beta = 0.5$ 
above the critical temperature (disordered phase), $J=1$, and for the $s$ values
$0.1$, $s_c = 0.28$, and $0.6$.
}
\label{figmaxhigh}
\end{figure}

\begin{figure}
\begin{center}
\includegraphics[height=50mm]{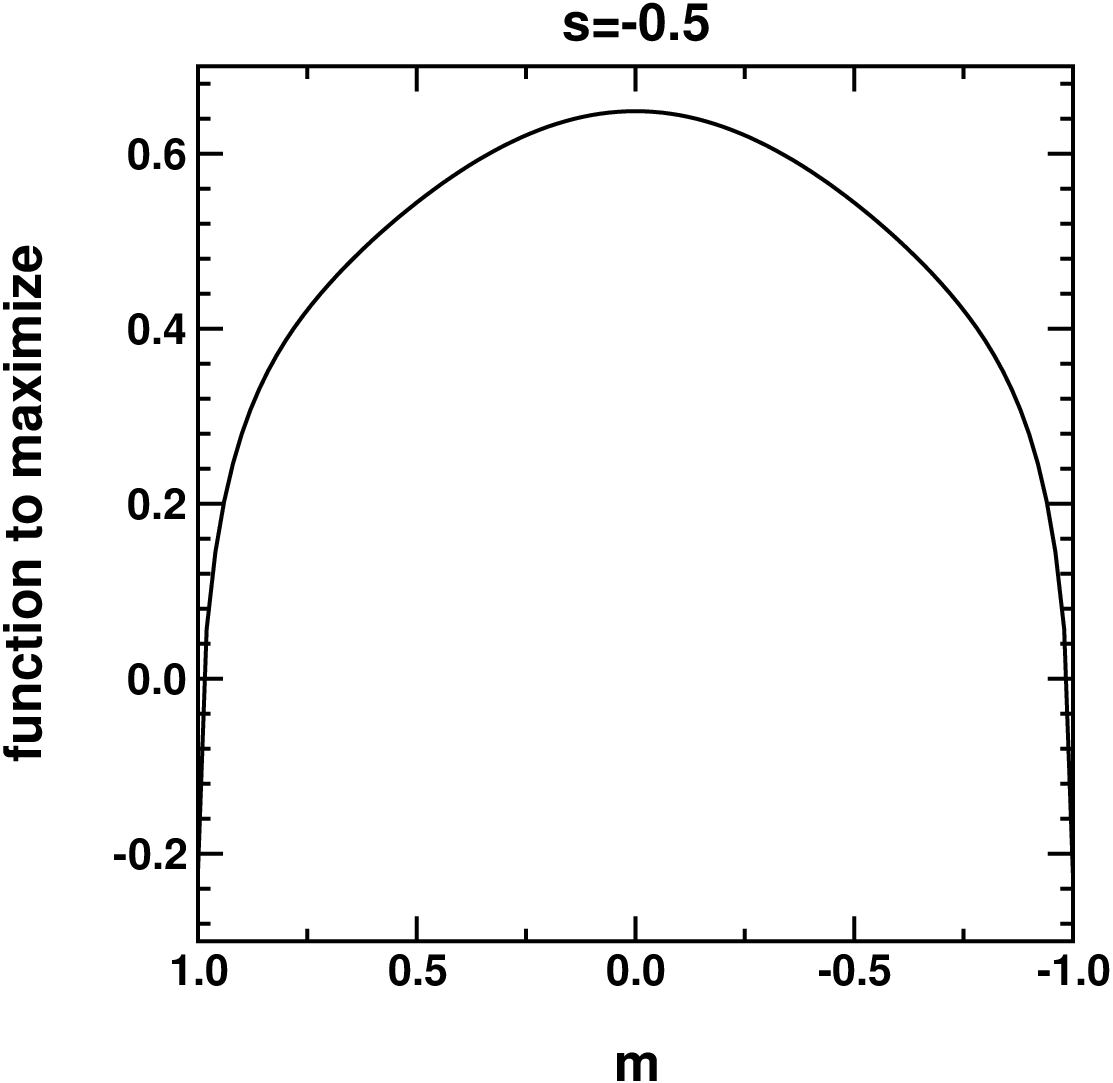}
\includegraphics[height=50mm]{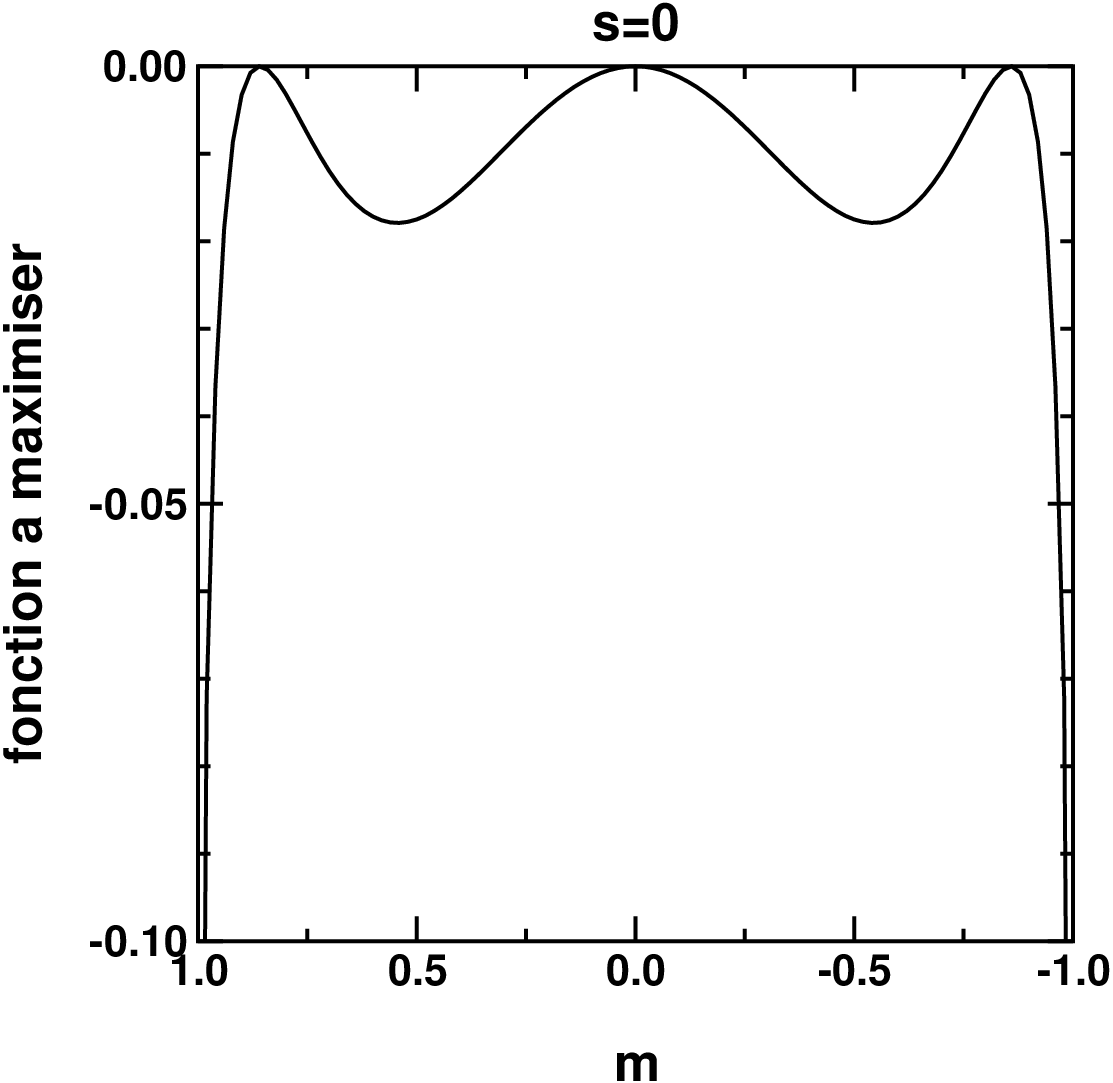}
\includegraphics[height=50mm]{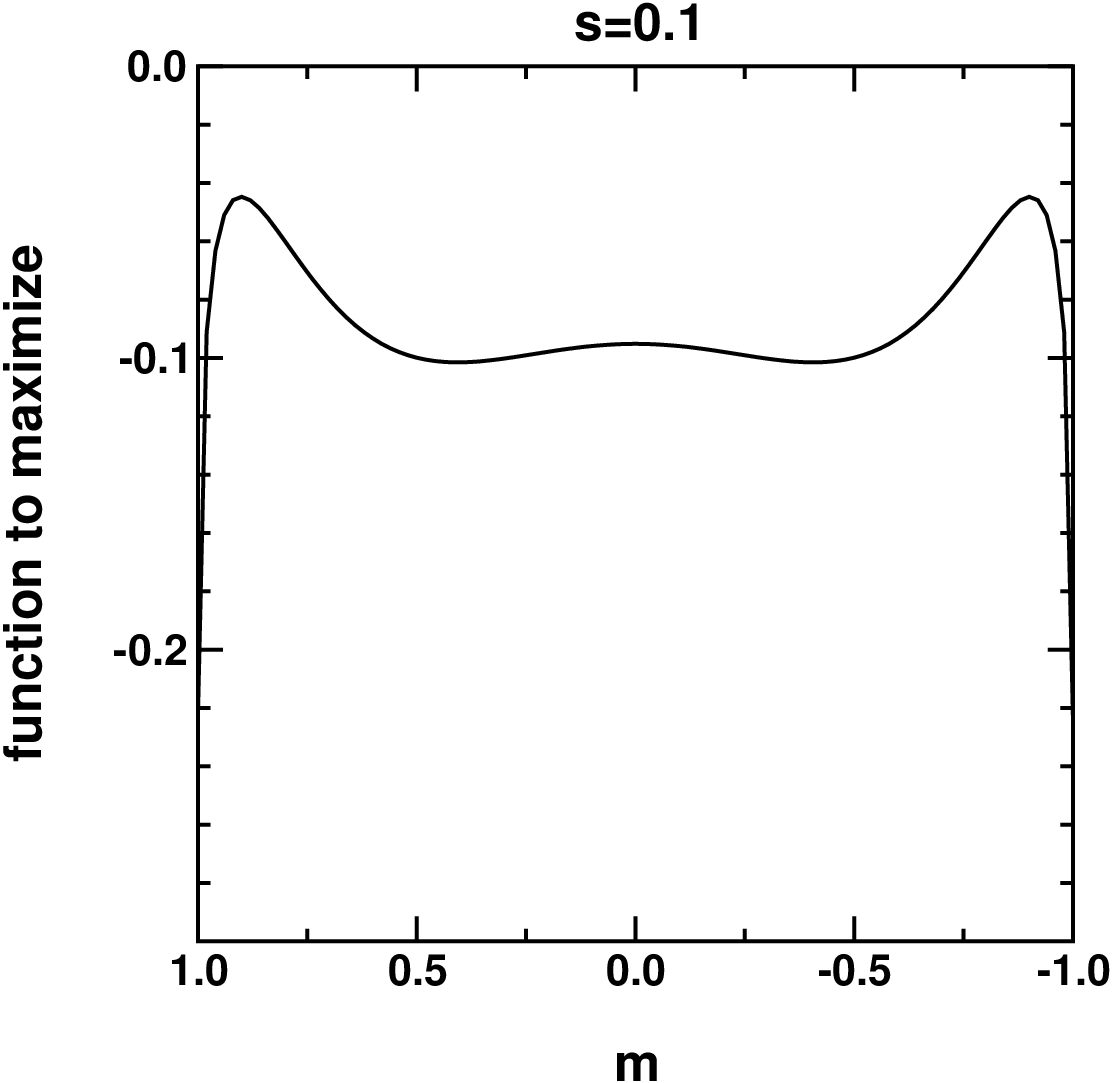}
\end{center}
\caption{The function that must be maximized in order to find
the value of $m_K(s)$ and $\frac{\psi_K(s)}{N}$, in the case
of the Ising model (q=2), for a temperature $\beta = 1.5$
under the critical temperature (ordered phase), $J=1$, and for the $s$ values
$-0.5$, $0$, and $0.1$.
}
\label{figmaxlow}
\end{figure}

\begin{figure}
\begin{center}
\includegraphics[height=50mm]{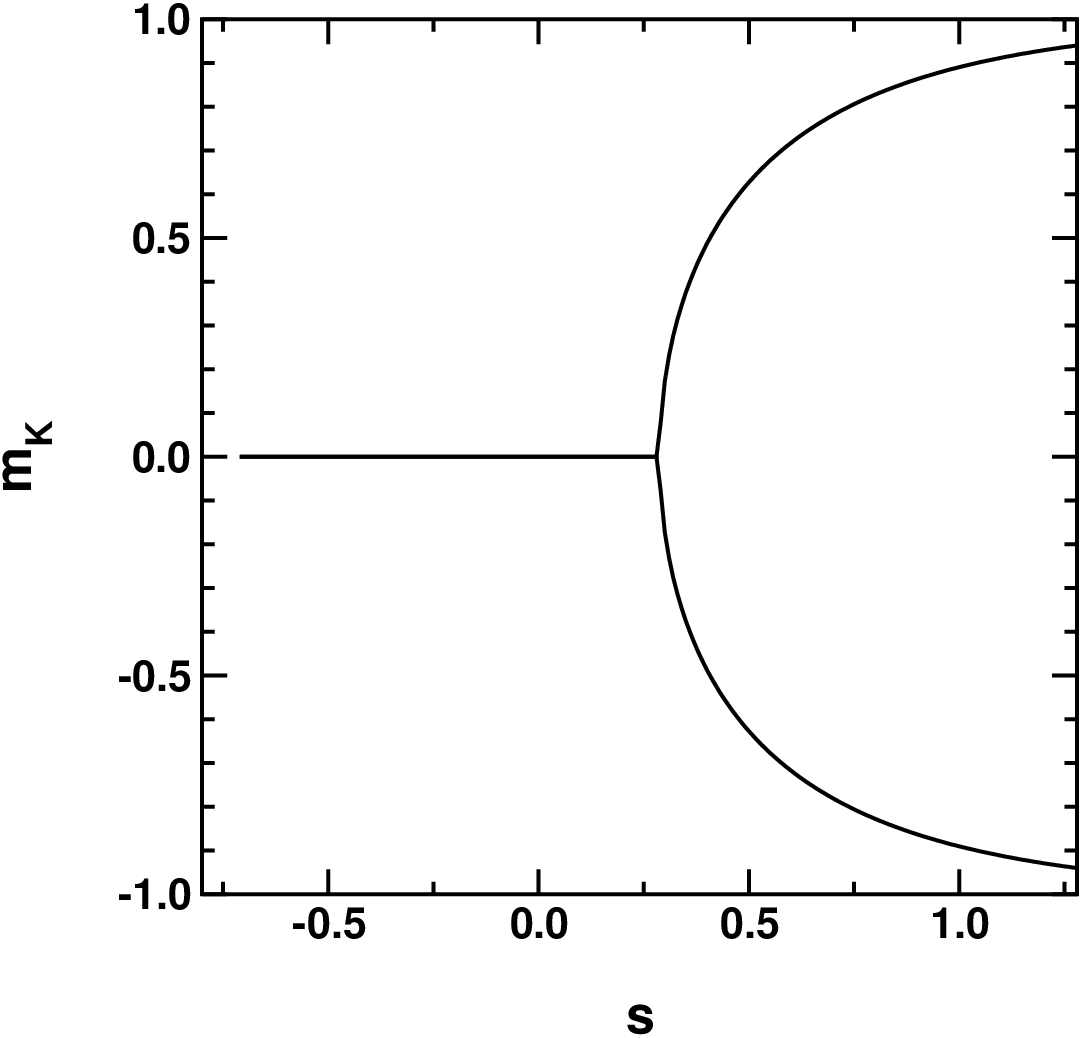}
\hskip 2cm
\includegraphics[height=50mm]{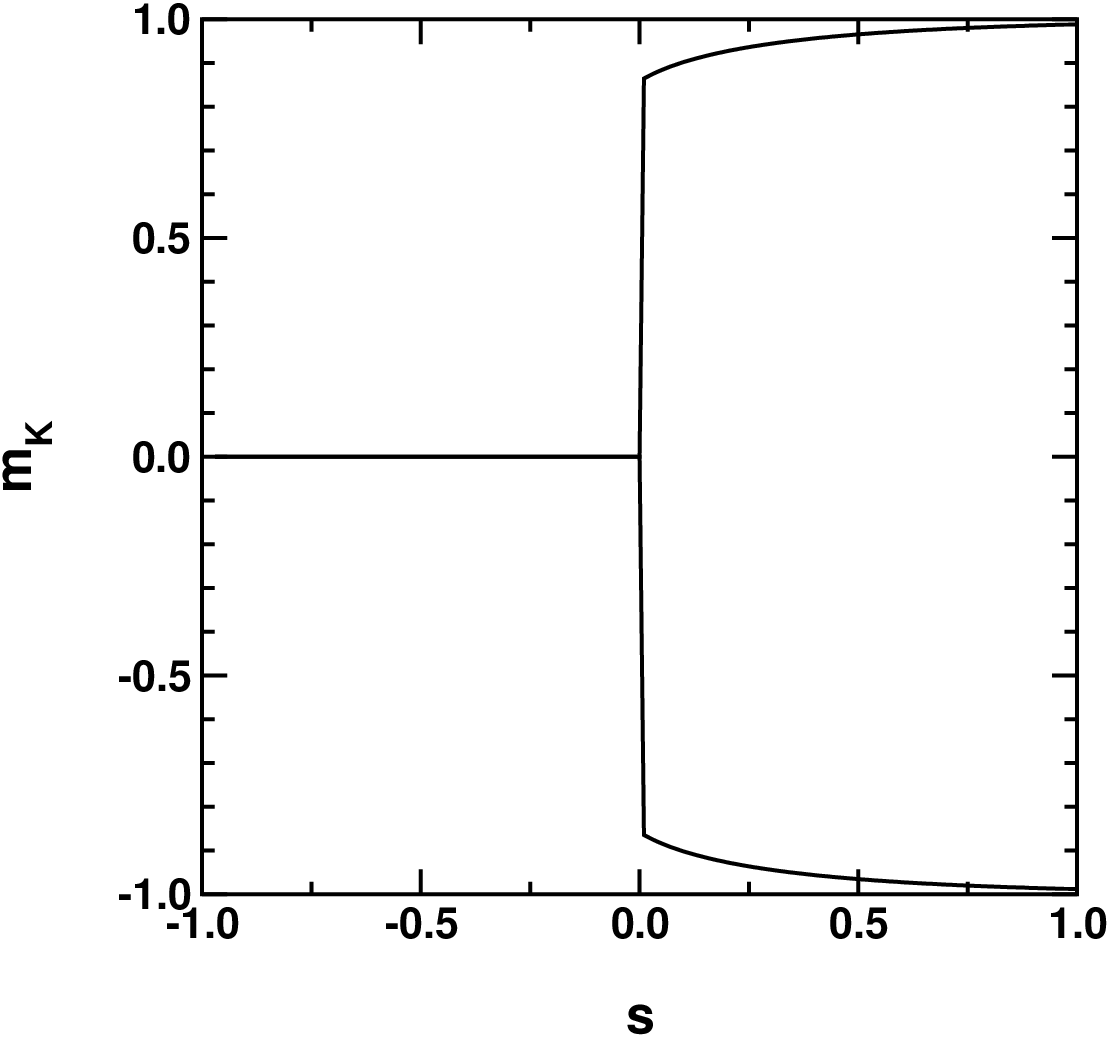}
\end{center}
\caption{The value of the magnetization for which the function of
(\ref{maxeqising}) is maximal, as a function of $s$,
in the case of the Ising model (q=2),
for $\beta=0.5$
(left) and $\beta=1.5$ (right). We took $J=1$.}
\label{figmkising}
\end{figure}

\begin{figure}
\begin{center} 
\includegraphics[height=60mm]{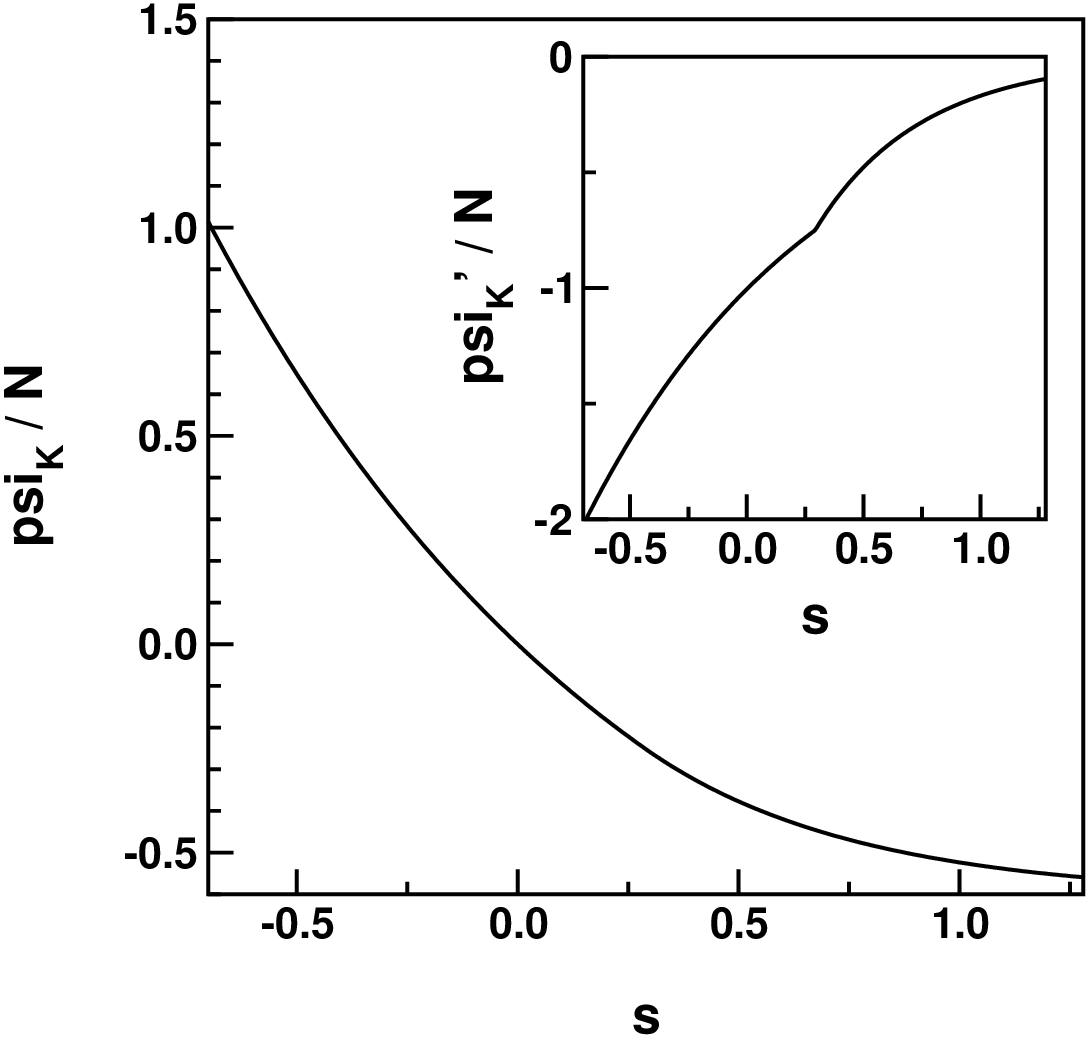}
\hskip 2cm
\includegraphics[height=60mm]{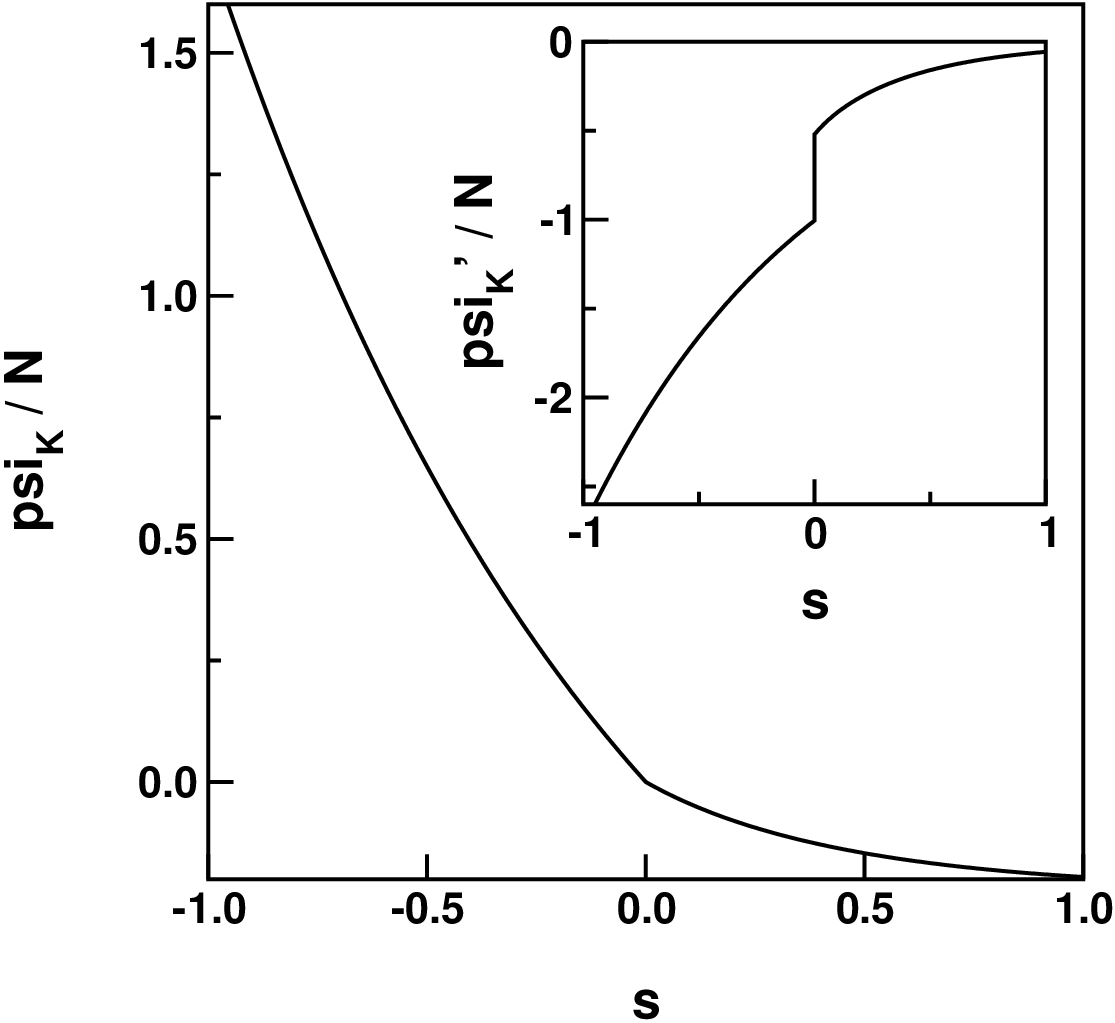}
\end{center}
\caption{$\psi_K(s)$ in the case of the Ising model (q=2), for $\beta=0.5$
(left) and $\beta=1.5$ (right). We took $J=1$. The derivative $\psi_K'(s)$
is plotted in the inset. It is continuous (resp. discontinuous) in the disordered (ordered) phase.}
\label{figpsikising}
\end{figure}

For $q=2$, spins can take only two values $S_i = \pm 1$. The corresponding
fractions can be expressed in terms of a unique order parameter,
the magnetization $m$:
\begin{equation*}
\theta_1 =\frac{1+m}{2} \;\; \mbox{ and } \;\; \theta_2 =\frac{1-m}{2}
\end{equation*}
with $\theta_1-\theta_2=m$, and the Hamiltonian reads with these
notations
\begin{equation}
\HH = -\frac{J}{2N} \sum_{i,j} S_i S_{j}.
\end{equation}
Then the maximization condition (\ref{maxeqq}) becomes
\begin{equation}
\frac{\psi_K(s)}{N} = \max_{m} \left\{
z \sqrt{(1-m^2)} - \cosh{\beta J m} + m \sinh{\beta J m}\right\}
\label{maxeqising}
\end{equation}
The behavior of this function is illustrated in figures \ref{figmaxhigh}
and \ref{figmaxlow}. The magnetization for which the maximum is
obtained is called $m_K(s)$.
In the high temperature phase, it is vanishing for a whole range of $s$
values. This was of course
expected for $s=0$, where $m_K$ coincides with the physical magnetization
of the system.
But above a certain threshold $s_c$, the maximum splits into two symmetrical
ones. The transition is continuous (see figure \ref{figmkising}, left),
and corresponds to a discontinuity in the second derivative of $\psi_K(s)$
(figure \ref{figpsikising}).
A clear difference appears between the disordered and ordered phases.
In the latter case ($\beta>1$), the transition value $s_c$ is exactly at zero
(figure \ref{figmkising}, right).
Besides, the transition in $m_K$ is now discontinuous, and the first derivative
of $\psi_K(s)$ already shows a discontinuity.


Once we know that the transition above $T_c$ is continuous,
and thus that $m_K(s)$ is close to zero at the transition,
it is possible to determine the value for $s_c$ by expanding
the expression for $\psi_K(s)$ for small $m_K$ values.
Then the condition $\frac{d \psi_K}{d m_K} = 0$ yields
a non-zero $m_K$ solution for
\begin{equation}
z_c = \beta J\left(2-\beta J\right)
\end{equation}

It is of course much harder to perform these calculations in finite dimension. It may nevertheless be argued that the dynamical phase transition occurring in the paramagnetic phase for
$s>s_c>0$
has an upper critical dimension $d_c=6$   characterized by critical exponents that are apparently unrelated to those of the static phase transition at $\beta=1$ and $s=0$.

\subsection{$q = 3$: the three states Potts model}

For $q=3$,
the critical temperature of the thermodynamic
phase transition corresponds to $\beta_c = 2.7725$ \cite{mendes_l91}.
As we shall show graphically, 
the dynamical phase transition is now discontinuous
both above and under the critical temperature

\begin{figure}
\begin{center}
\includegraphics[height=35mm]{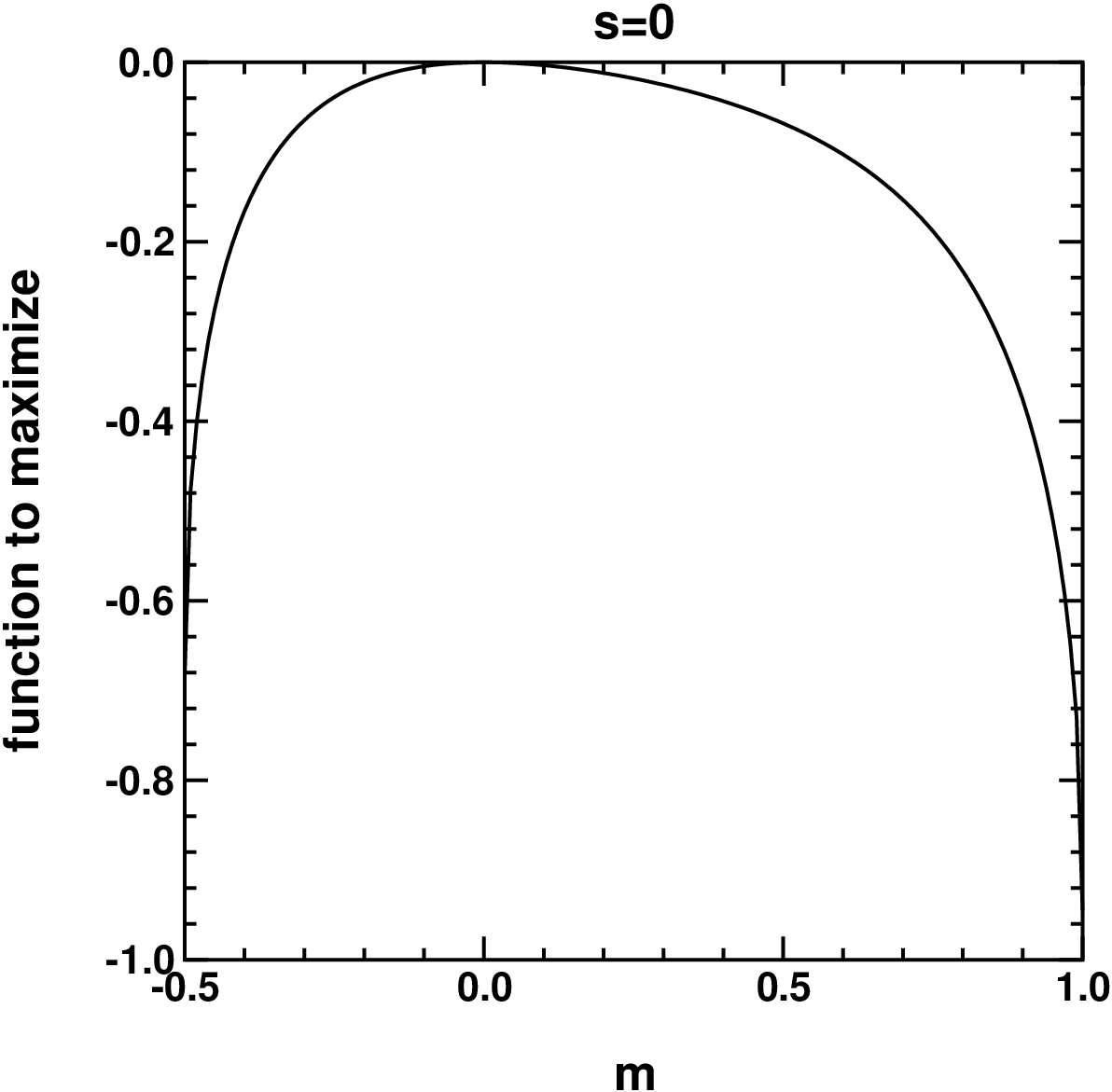}
\includegraphics[height=35mm]{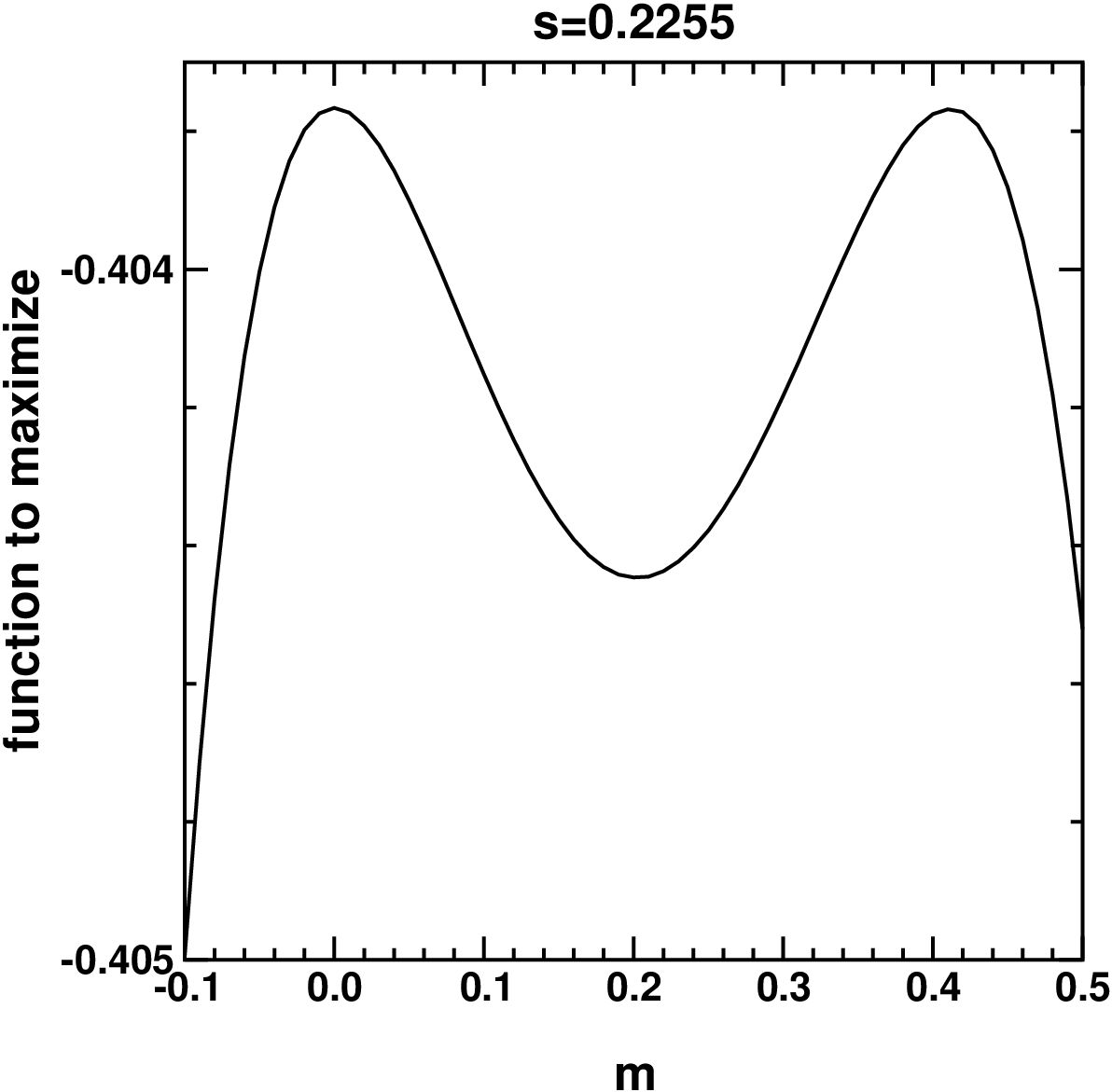}
\includegraphics[height=35mm]{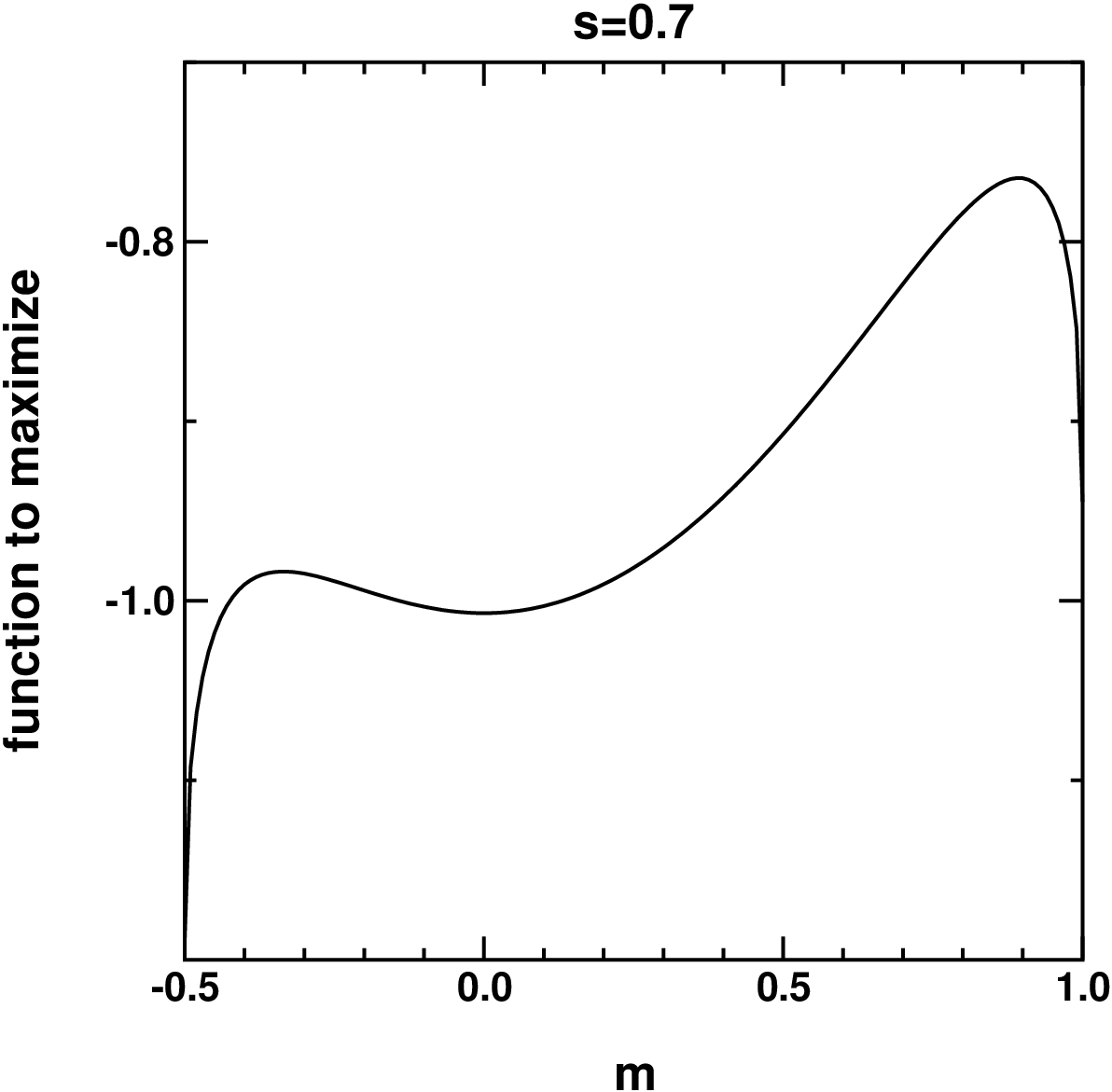}
\includegraphics[height=35mm]{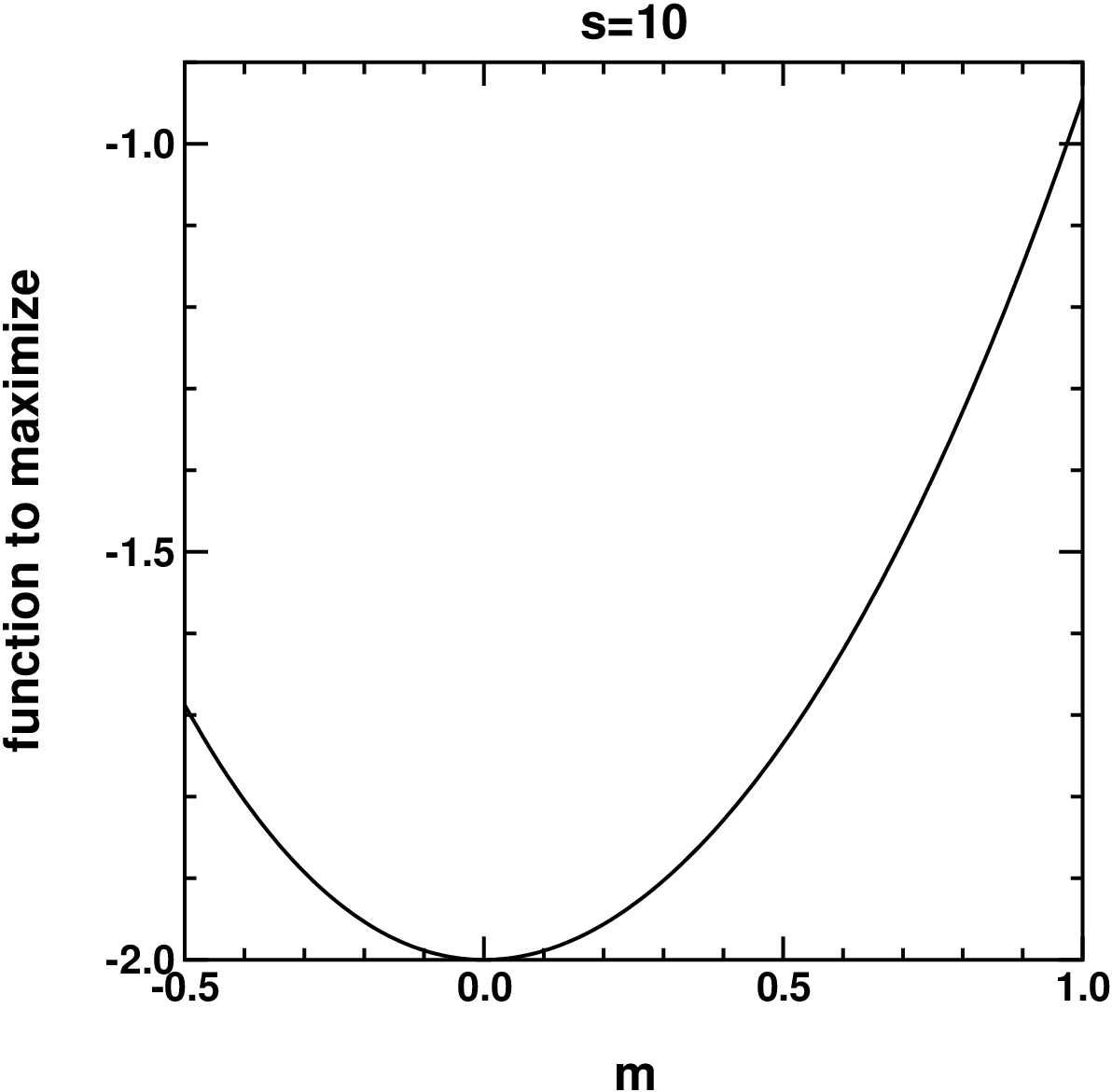}
\end{center}
\caption{The function that must be maximized in order to find
the value of $m_K(s)$ and $\frac{\psi_K(s)}{N}$, in the case
of the Potts model for $q=3$, for a temperature
$\beta = 0.5$ 
above the critical temperature, for $J=1$, and for the $s$ values
$0$, $s_c = 0.2255$, $0.7$ and $10$.
}
\label{figmax3high}
\end{figure}

\begin{figure}
\begin{center}
\includegraphics[height=40mm]{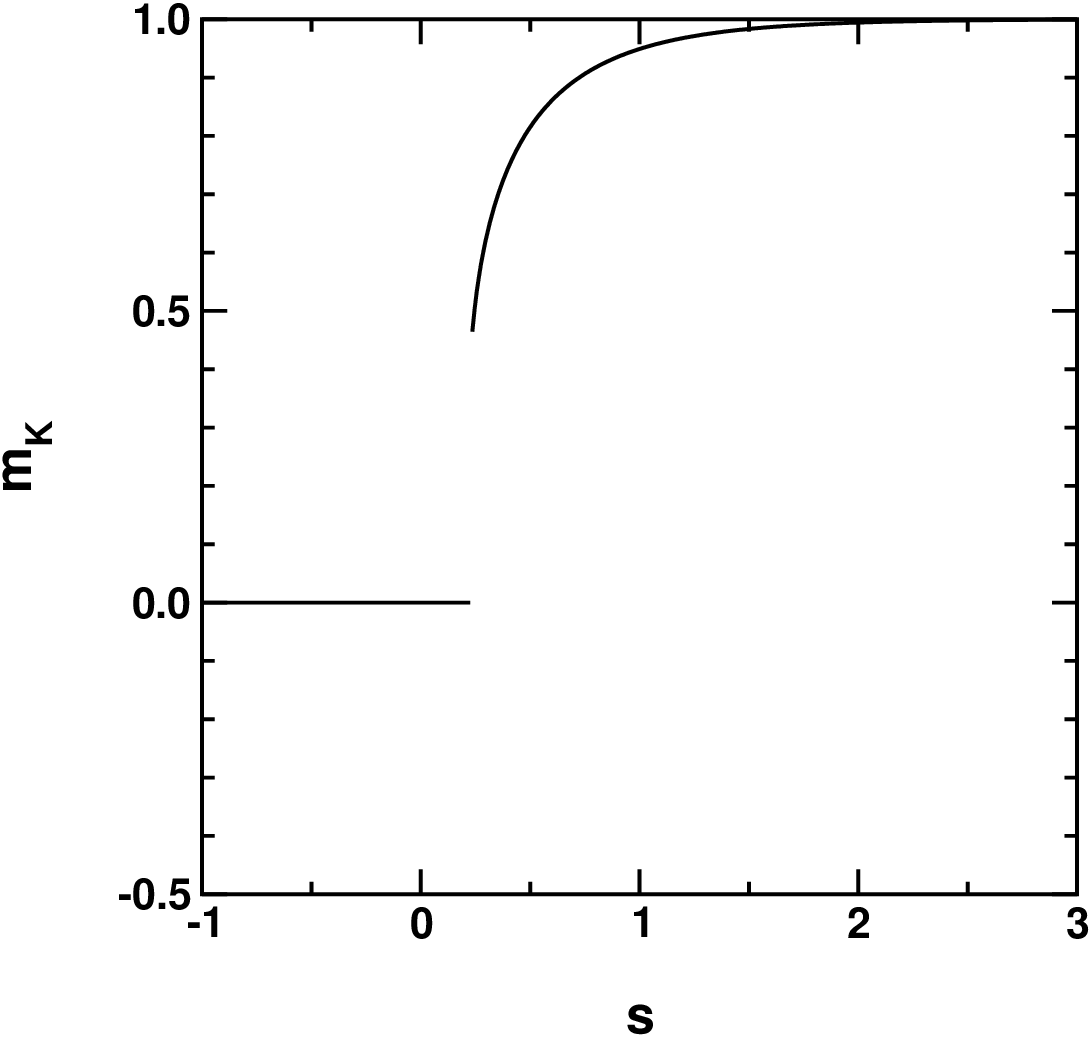}
\hskip 2cm
\includegraphics[height=40mm]{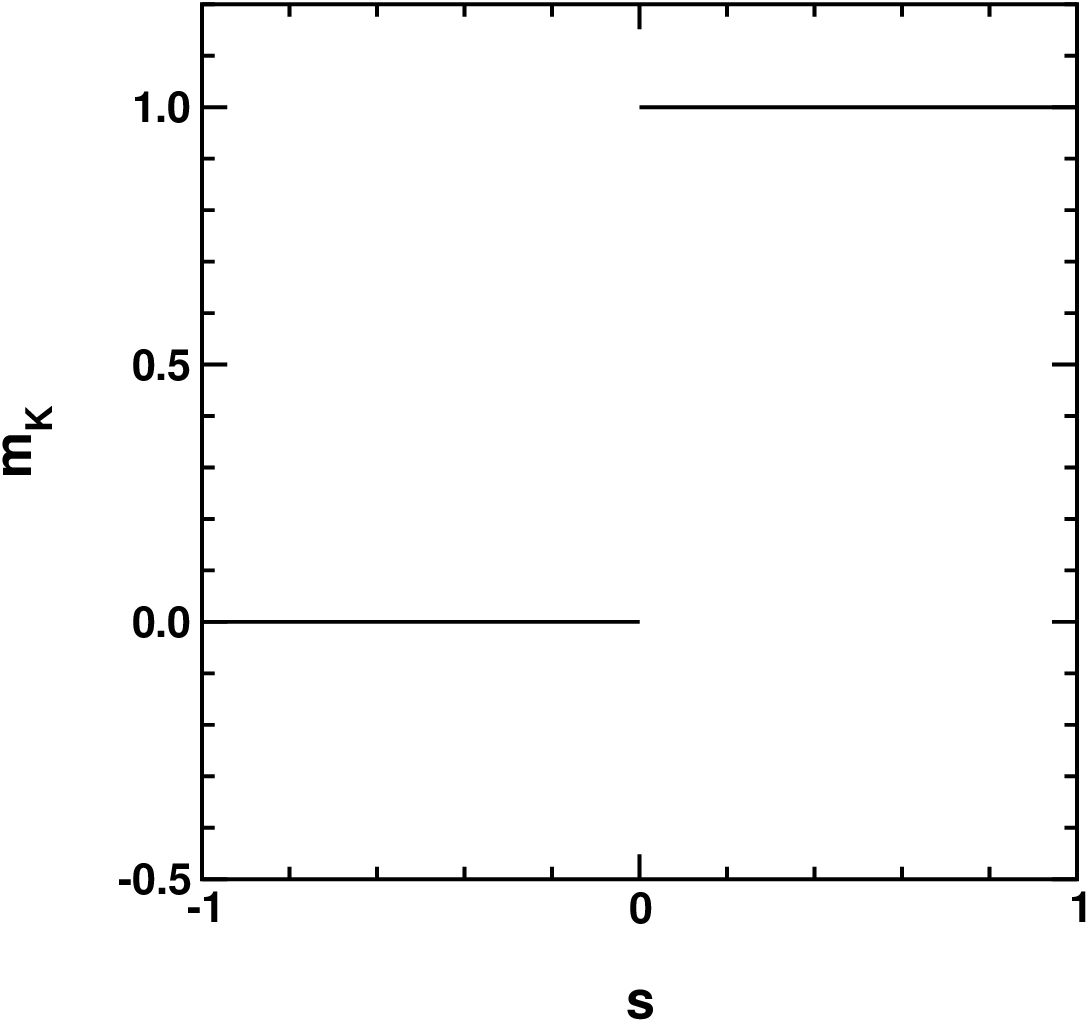}
\end{center}
\caption{The value of the magnetization for which the function of
(\ref{maxeqising}) is maximal as a function of $s$, in the case
of the Potts model for $q=3$, for $\beta=0.5$
(left) and $\beta=5$ (right). We took $J=1$.}
\label{fig_q3m}
\end{figure}

\begin{figure}
\begin{center} 
\includegraphics[height=60mm]{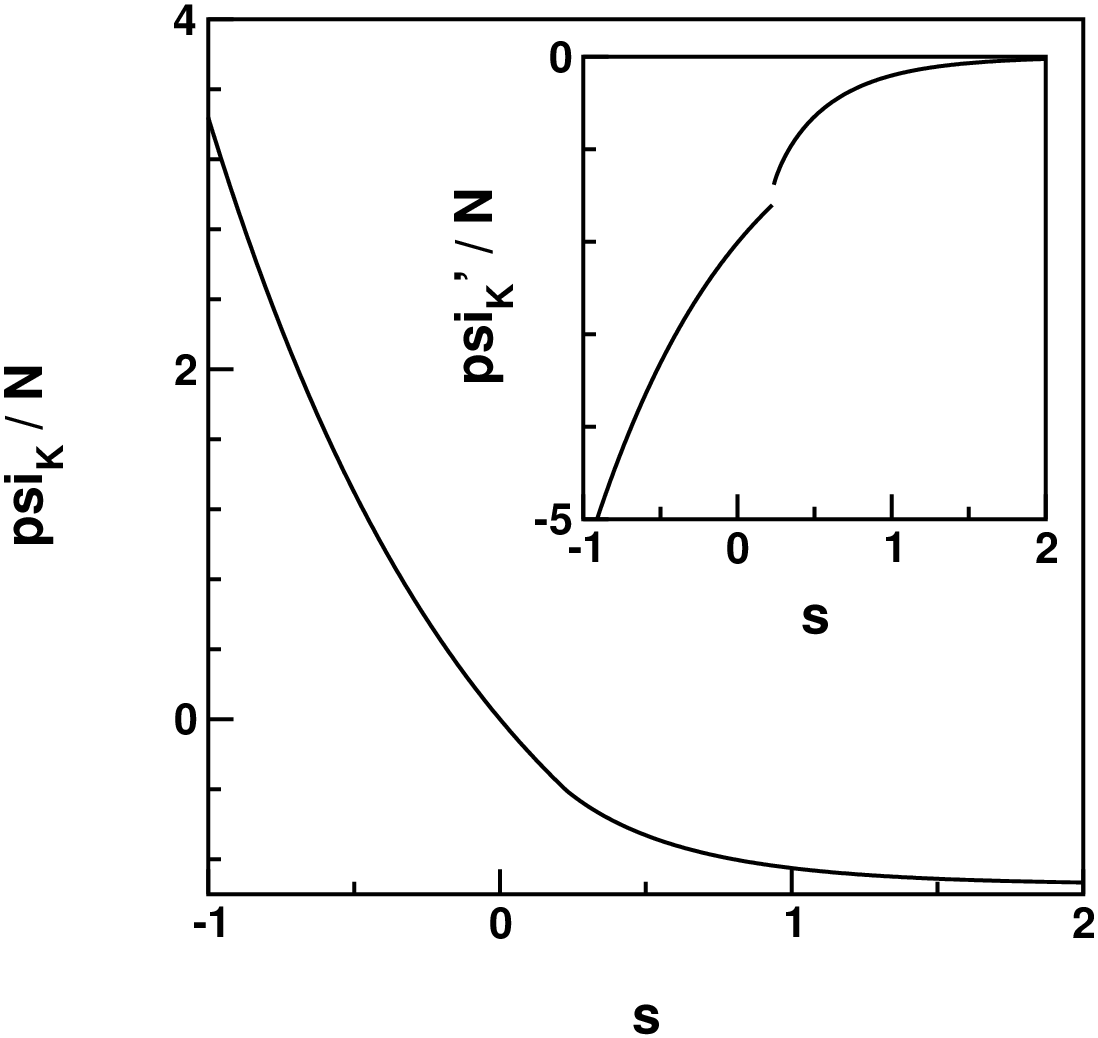}
\hskip 2cm
\includegraphics[height=60mm]{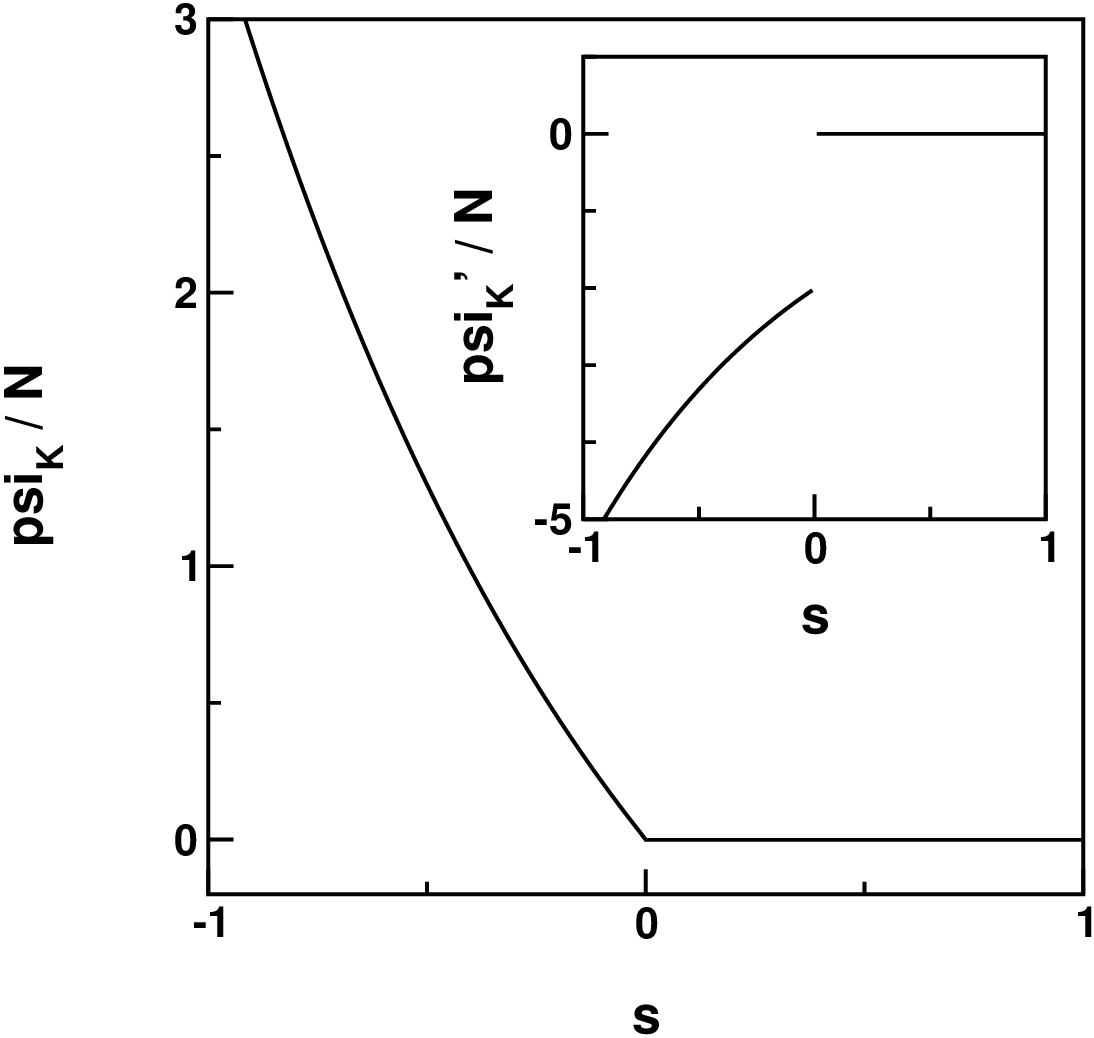}
\end{center}
\caption{$\psi_K(s)$, in the case
of the Potts model for $q=3$, for $\beta=0.5$
(left) and $\beta=5$ (right). We took $J=1$. The derivative $\psi_K'(s)$
is plotted in the inset. It is discontinuous in both cases.}
\label{fig_q3psik}
\end{figure}

For $\beta=0.5$, figure \ref{figmax3high} shows the behavior
of the function to maximize.
By contrast with the Ising case, now the transition in the $m_K$ parameter
is discontinuous even in the high temperature phase (figure \ref{fig_q3m}).
Both for high and low temperatures, the discontinuity occurs
in the first derivative of $\psi_K(s)$, as seen in figure \ref{fig_q3psik}.
However, while the transition at high temperature occurs at $s=s_c \neq 0$,
it occurs at $s=0$ under the critical temperature.


\subsection{The $q=1$ limit}

The limit $q=1$ is interesting because it can be related
to percolation (see \cite{lubensky79} for example).
In order to have a well defined limit, one considers
\begin{equation*}
\tilde{\psi}_K(s) = \lim_{q\to 1} \frac{\psi_K(s)}{q-1} =
N \max_{m \leq 1} \left\{ z \left( 2\sqrt{1-m} + m-1\right)
+ (1-m)\left( 1- \ee^{\frac{\beta J}{2} m} \right)
- \ee^{- \frac{\beta J}{2} m}
\right\}
\end{equation*}
In the case $z=1$, this expression can be factorized
\begin{equation}
\frac{\tilde{\psi}_K(s)}{N}
= \max_{m \leq 1} \left\{ \sqrt{1-m} \ee^{\frac{\beta J}{4} m}
- \ee^{- \frac{\beta J}{4} m}
\right\}^2
\end{equation}
As in the case $q=3$, one finds that the dynamical phase transition
that occurs at $s=0$ for $\beta>1$ or $<1$ is characterized by a discontinuous
$m_K(s)$, and a continuous cumulant generating function (the discontinuity occurs
in the first derivative of $\psi_K(s)$).

\section{Conclusion}

Though the thermodynamic formalism was designed in order to study
out-of-equilibrium systems, it is also a useful tool to shed a new light
on equilibrium systems. Their dynamical properties turn out to be richer
than could be expected. Actually, some dynamical phase transitions
do occur even in apparently dull phases as the disordered phase in the
Ising model. Dynamical phase transitions and thermodynamic ones are
connected : in all examples that we studied, the features of the dynamical
phase transition are different on each side of the thermodynamic phase
transition. The understanding of the interplay between dynamical and
thermodynamic phase transitions could lead to a renewed
vision of equilibrium systems.

Here the case of the Potts model was studied for a various number of states
$q$.
Interestingly, the case $q=2$ turns out to be quite special in the sense
that the paramagnetic phase of the Ising model is the only one where
a continuous phase transition occurs in the order parameter $m_K(s)$, corresponding to
a discontinuity which appears only in the second derivative of $\psi_K(s)$.
By contrast, both for $q=1$ or $q=3$, all dynamical transitions
were found to be discontinuous in the order parameter $m_K(s)$, 
and it is the first derivative of $\psi_K(s)$ that shows a discontinuity.

In a research field which often requires very technical and heavy
calculations, the method that we present here is interestingly simple.
We applied it in this paper to the calculation of the cumulant
generating function of $K$ for the mean-field Potts model in the large
$N$ limit, but more
generally it can be applied
to any mean-field model which can be described by a macroscopic order
parameter, and for which a detailed balance relation at the macroscopic
scale is available (in particular, the transition rates of the dynamics
must be functions only of the aforementionned order parameter).
Though all the results we presented were dealing with the observable 
$K$, the same kind of calculations - with the same kind of conclusions 
in the case of the Potts model -
can be performed for the observable $Q_+$ (the corresponding
cumulant generating function is then the so-called topological pressure).
It should be noticed however that in the case of $Q_+$, the large $N$
expansion of equation (\ref{eq_dvlpt}) is only valid for $s<1$, as a consequence
of $1-s$ powers.
A generalization to other observables of the form
(\ref{defA}) with $\alpha(\C,\C') = \alpha(\bN,\bN ')$ is often possible. 





\begin{thebibliography}{10}

\bibitem{ruelle78}
David Ruelle.
\newblock {\em Thermodynamic Formalism}.
\newblock Addison Wesley Publ. Co. (Reading, Mass, 1978)., 1978.

\bibitem{gaspard04}
P.~Gaspard.
\newblock Time-reversed dynamical entropy and irreversibility in markovian
  random processes.
\newblock {\em J. Stat. Phys.}, 117:599--615, 2004.

\bibitem{vanbeijeren_d02}
H.~van Beijeren and J.R. Dorfman.
\newblock A note on the {R}uelle pressure for a dilute disordered {S}inai
  billiard.
\newblock {\em J. Stat. Phys.}, 108:516, 2002.

\bibitem{latz_b_d96}
A.~Latz, H.~van Beijeren, and J.R. Dorfman.
\newblock Lyapunov spectrum and the conjugate pairing rule for a thermostated
  random {L}orentz gas: kinetic theory.
\newblock {\em Phys. Rev. Lett.}, 78:207--210, 1997.

\bibitem{lecomte_a_v05}
V.~Lecomte, C.~Appert-Rolland, and F.~van Wijland.
\newblock Chaotic Properties of Systems with Markov Dynamics
\newblock {\em Phys. Rev. Lett}, 95:010601, 2006.

\bibitem{lecomte_a_v06}
V.~Lecomte, C.~Appert-Rolland, and F.~van Wijland.
\newblock Thermodynamic formalism for systems with {M}arkov dynamics.
\newblock {\em To appear in J. Stat. Phys}, 2007.

\bibitem{gaspard98}
P.~Gaspard.
\newblock {\em Chaos, scattering and statistical mechanics}.
\newblock Cambridge Nonlinear Science Series vol. 9 (Cambridge UP, 1998), 1998.

\bibitem{lebowitz_s99}
J.L. Lebowitz and H.~Spohn.
\newblock A {G}allavotti-{C}ohen type symmetry in the large deviation
  functional for stochastic dynamics.
\newblock {\em J. Stat. Phys.}, 95:333--365, 1999.

\bibitem{derrida_a99}
B.~Derrida and C.~Appert.
\newblock Universal large deviation function of the {K}ardar-{P}arisi-{Z}hang
  equation in one dimension.
\newblock {\em J. Stat. Phys.}, 94:1--30, 1999.

\bibitem{mendes_l91}
J.F.F. Mendes and E.J.S. Lage.
\newblock Dynamics of the infinite ranged {P}otts model.
\newblock {\em J. Stat. Phys.}, 64:653, 1991.


\bibitem{lubensky79}
T.C. Lubensky.
\newblock In Maynard Balian and Toulouse, editors, {\em Ill Condensed Matter,
  Les Houches XXXI}. North-Holland, Amsterdam, 1979.

\end{thebibliography}



\end{document}